\documentclass[conference]{IEEEtran}

\usepackage{tabularx}
\usepackage{booktabs}
\usepackage[pdftex]{graphicx}
\let\labelindent\relax
\usepackage{enumitem}
\usepackage{multirow}
\usepackage{amsmath}
\usepackage[multiple]{footmisc}
\usepackage{mdframed}
\usepackage[linesnumbered,ruled]{algorithm2e}
\usepackage{algpseudocode}
\usepackage{makecell}

\SetAlCapNameFnt{\footnotesize}
\SetAlCapFnt{\footnotesize}
\let\oldnl\nl
\newcommand{\nonl}{\renewcommand{\nl}{\let\nl\oldnl}}

\begin{document}

\title{Extracting and Analyzing Context Information\\ in User-Support Conversations on Twitter}

\author{
    \IEEEauthorblockN{Daniel Martens}
    \IEEEauthorblockA{
        University of Hamburg\\
        Hamburg, Germany\\
        martens@informatik.uni-hamburg.de
    }
    \and
    \IEEEauthorblockN{Walid Maalej}
    \IEEEauthorblockA{
        University of Hamburg\\
        Hamburg, Germany\\
        maalej@informatik.uni-hamburg.de
    }
}

\maketitle
\begin{abstract}
While many apps include built-in options to report bugs or request features, users still provide an increasing amount of feedback via social media, like Twitter. Compared to traditional issue trackers, the reporting process in social media is unstructured and the feedback often lacks basic context information, such as the app version or the device concerned when experiencing the issue. To make this feedback actionable to developers, support teams engage in recurring, effortful conversations with app users to clarify missing context items.

This paper introduces a simple approach that accurately extracts basic context information from unstructured, informal user feedback on mobile apps, including the platform, device, app version, and system version. Evaluated against a truthset of 3014 tweets from official Twitter support accounts of the 3 popular apps Netflix, Snapchat, and Spotify, our approach achieved precisions from 81\% to 99\% and recalls from 86\% to 98\% for the different context item types.
Combined with a chatbot that automatically requests missing context items from reporting users, our approach aims at auto-populating issue trackers with structured bug reports.
\end{abstract}


\IEEEpeerreviewmaketitle


\section{Introduction}

Modern apps include options to support users in providing relevant, complete, and correct context information when reporting bugs or feature requests. For example, Facebook attaches more than 30 context items to bug reports submitted via their apps, including the app version installed and the device in use \cite{Weblink:100}. Despite the presence of such options, an increasing amount of users still report their issues via social media, such as Twitter. A possible reason might be to increase the pressure on software vendors through the public visibility of reported issues. Research has shown, that mining tweets allows additional features and bugs to be extracted, that are not reported in official channels as app stores~\cite{Nayebi:2018:EMSE}. Mezouar et al. found that one third of the bugs reported in issue trackers can be discovered earlier by analyzing tweets~\cite{Mezouar:2018:TUB:3231288.3231333}. Many app vendors are aware of these benefits and have thus created Twitter support accounts as \textit{@Netflixhelps}, \textit{@Snapchatsupport}, or \textit{@SpotifyCares}.

\begin{figure}[]
\centering
\includegraphics[width=.95\columnwidth]{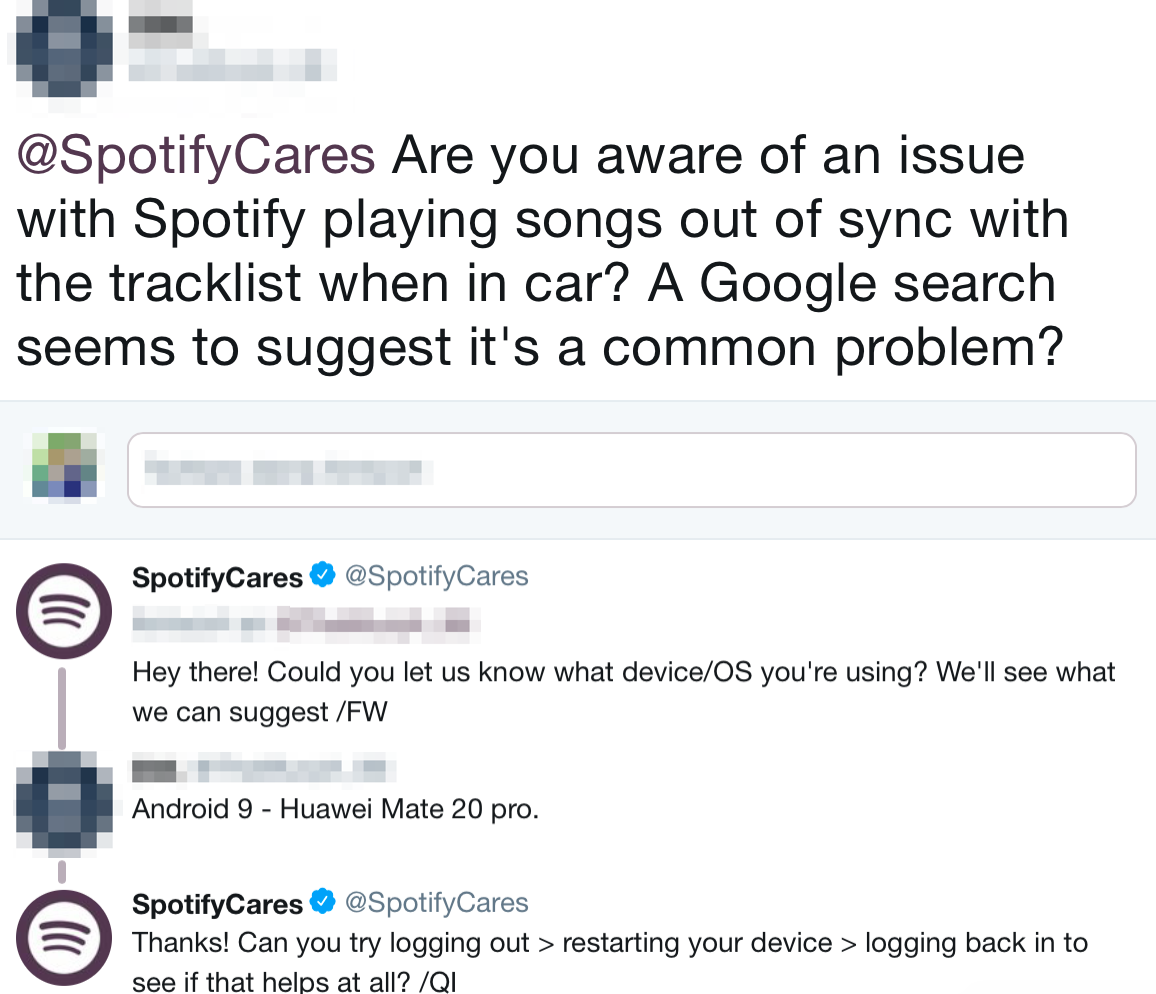}
\caption{Example of a Conversation between User and Support Team on Twitter to obtain Missing Basic Context Items (i.e., the Device and Platform).}
\label{fig:twitter-conversation}
\end{figure}

Compared to structured reports in issue trackers that usually include context items \cite{bettenburg-fse-2008, bettenburg-msr-2008}, feedback on Twitter is primarily provided by non-technical users in a less structured way~\cite{Mezouar:2018:TUB:3231288.3231333}. 
Tweets that miss basic context items, such as the concerned platform, are likely to be non-actionable to developers. Hence, several support accounts prominently highlight the importance  of this information in their Twitter bio. For instance, Spotify's profile includes \textit{``for tech queries, let us know your device/operating system''}, while Netflix states \textit{``for tech issues, please include device \& error''}.
However, tweets, such as 

\begin{quote}
\textit{``I can't open playlists shared via WhatsApp on my iPhone XR, iOS 12.1.4, Spotify 8.4.61''}
\end{quote}

\noindent
that include all basic context items, i.e., the concerned platform, device, app and system version, are rare.
In contrast, support teams engage in recurring, effortful conversations with users to obtain missing information, as shown in Figure~\ref{fig:twitter-conversation}.

The overall goal of our research is to support both users and developers in exchanging precise context information with the least possible effort. As a first step to this end, this paper discusses the automatic identification of context items in informal user-support conversions related to mobile apps.
We introduce a simple unsupervised approach that uses pre-defined keyword lists, word vector representations, and text patterns to extract basic context items from tweets, including the platform, device, app version, and system version. The results allow to identify issues potentially actionable to developers or requiring further clarifications. Evaluated against a truthset of $\sim$3,000 tweets, our approach achieved precisions from 81\% to 99\% and recalls from 86\% to 98\% for the different context item types.

\begin{figure*}[]
\centering
\includegraphics[width=.85\textwidth]{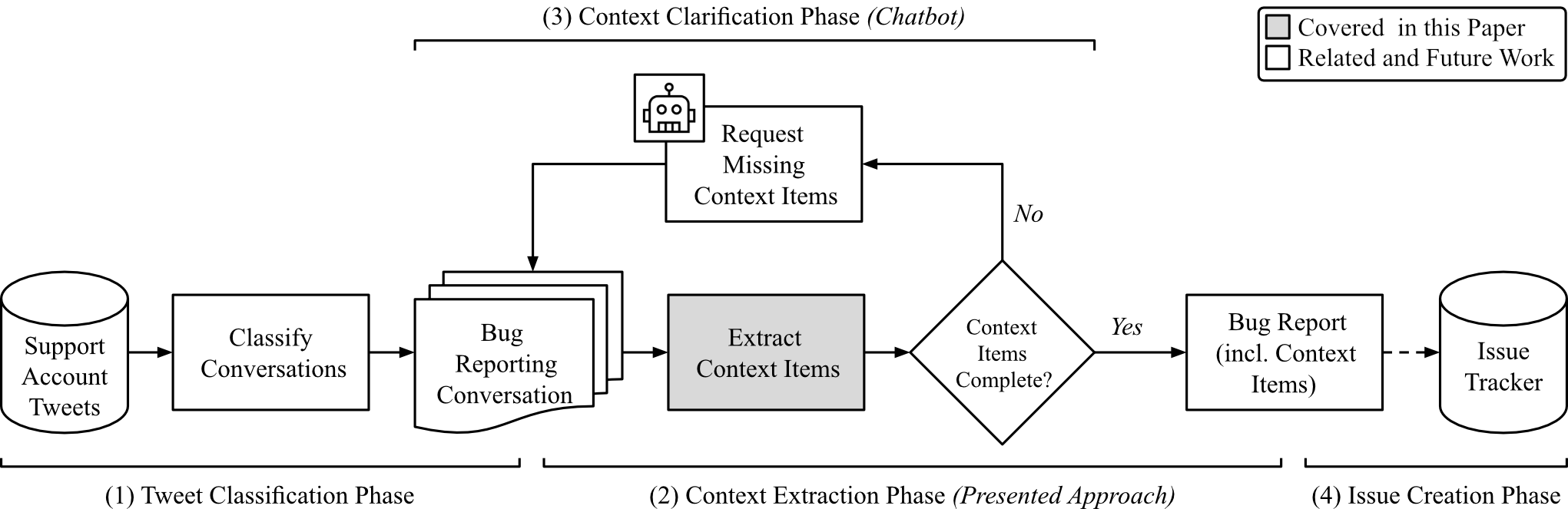}
\caption{Overall Setting to Auto-Populate Issue Trackers with Structured Bug Reports Including Context Items mined from User Tweets.}
\label{fig:approach-chatbot}
\end{figure*}

The remainder of the paper is structured as follows: Section~II describes our research setting. Then, Section~III introduces our approach to extract context items and Section~IV reports on the evaluation results. Section~V discusses the findings and potential threats to validity. Finally, Section~VI surveys related work and Section~VII concludes the paper.


\section{Research Setting}
We describe the overall usage setting for our context extraction approach as well as our research method and data.


\subsection{Overall Setting of this Work}

Developers organize their work using issue trackers~\cite{6698918}.
An issue usually corresponds to a unit of work to accomplish an improvement in a software system.
Issues can be of different types, such as bug reports or feature requests.
When creating an issue of a specific type, issue trackers use structured templates that request specific context items to be provided by the reporter.
Bug reports require, e.g., the affected app version, while feature requests require a description of the desired feature.
Traditionally, reporters were technically experienced persons, such as software testers or the developers themselves.

With the emergence of app stores and social media, also non-technical users began to frequently and informally communicate with developers -- compared to existing public issue trackers of open source projects that were tailored towards more technical experienced users. 
Research has shown that users include requirements-related information such as bug reports in about one third of their informal feedback~\cite{6224306, 6636712, 7320414}. 
Recent studies specifically emphasized the benefits of mining  tweets~\cite{8048893, 8048885}.

There are several key challenges software practitioners face when working with bug reports included in informal user feedback, e.g. provided via app stores and social media:

\begin{enumerate}[label=\textit{(\arabic*)}, leftmargin=*]
  \item \textit{Missing Information.} Compared to reports in issue trackers, feedback in app stores and social media is primarily provided by non-technical users in a less structured way~\cite{Mezouar:2018:TUB:3231288.3231333}. Unfortunately, users often miss to provide context items needed by developers such as the app version  ~\cite{bettenburg-fse-2008, 6636712, Joorabchi:2014kn, Breu:2010hi, Moran:2015hh}. This is compounded by online review processes that are purposefully unguided~\cite{Seyff2014} and lack quality checks, to allow many users to participate.

  \item \textit{Unreproducible Issues.} In case user feedback that reports bugs misses relevant context information, these bugs might become hard to reproduce~\cite{Joorabchi:2014kn, Maalej:2011ku}. Even if developers are able to guess the user's interactions, an issue might only occur on specific combinations of device model and system version~\cite{Gomez:2016jp}. Research found that developers fail to identify erroneous configurations already for a low number of features~\cite{Melo:2016:DVA:2884781.2884831}.
  
  \item \textit{Manual Efforts.} For developers to be able to understand and reproduce reported issues, support teams engage in effortful conversations with users \cite{Hassan2018}. Within our crawled dataset including tweets from the Netflix, Snapchat, and Spotify support accounts, more than 40\% ($\sim$2.2 million) of the tweets are provided by support teams, possibly to clarify missing context items.
\end{enumerate}

We aim to automatically extract basic context items from tweets. Our approach is intended to be used in combination with a feedback classification and a chatbot approach to auto-populate issue trackers with structured bug reports mined from user feedback, as shown in Figure~\ref{fig:approach-chatbot}.
The overall setting can continuously be applied, e.g., to an app's Twitter support account. It can be separated into \textit{four phases}, of which the second phase is covered by this paper, while the remaining phases are left for future work. We briefly describe each of the phases in the following:

\begin{enumerate}[label=\textit{(\arabic*)}, leftmargin=*]
  \item \textit{Tweet Classification Phase.} In the first phase, tweets addressed to the app's support account are classified by their types of requirements-related information. Only tweets reporting bugs (i.e., issues that potentially require context items to be understandable and reproducible by developers), are passed to the next phase. Tweets including other types of information, such as praise (e.g., \textit{``This is the greatest app I've ever used.''}), are excluded from further analysis. These do not require context items and a chatbot requesting such information would annoy app users.

  \item \textit{Context Extraction Phase.} In this phase, our context extraction approach is applied to single tweets or conversations consisting of multiple tweets that report bugs. Each tweet is mined to extract the four basic context items, including the platform, the device, the app version, and the system version. For example, the tweet \textit{``The app widget has died and is now a rectangular black hole. Xperia xz3 running Android''}, includes the device and platform.
 After processing a complete conversation, the approach verifies if all four items could be extracted.
  
  \item \textit{Context Clarification Phase.} If the four basic context items could not be extracted, a chatbot requests the missing information. In case of the example above, the chatbot would request the app version and system version by replying to the tweet: \textit{``Hey, help's here! Can you let us know the app version you're running, as well as the system version installed? We'll see what we can suggest''}. The conversations are periodically analyzed to see if the user provided the missing context items.
  
  \item \textit{Issue Creation Phase.} Once all context items are present, they are used to create a structured bug report within the app's issue tracker. The comment section of the issue tracker remains connected with the conversation on Twitter, so that developers can directly communicate with the reporting user to ask for further clarification or inform the user once the issue is fixed.
\end{enumerate}

By automatically requesting missing context items, our approach reduces the manual effort for support teams, and aids developers by addressing the aforementioned challenges to facilitate actionable bug reports.


\begin{figure}[b]
\centering
\includegraphics[width=.95\columnwidth]{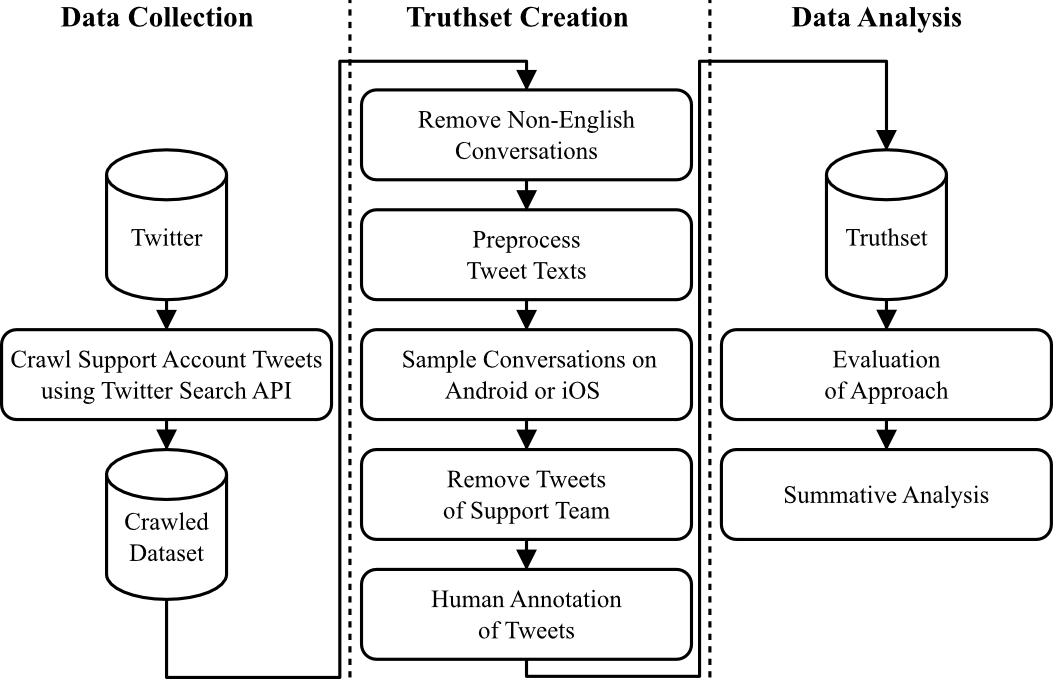}
\caption{Overview of the Research Method including the Data Collection, Data Preparation, Truthset Creation, and Data Analysis Phase.}
\label{fig:data-collection}
\end{figure}

\subsection{Research Method and Data}

In the following, we describe our research method including the data collection, truthset creation, and data analysis phase, as shown in Figure~\ref{fig:data-collection}.

\subsubsection{Data Collection Phase}

In the data collection phase, we crawled tweets using the Twitter Search API~\cite{Weblink:114} in January 2019. We refer to this data as \textit{crawled} dataset.

For our study, we collected tweets of the official Netflix, Snapchat, and Spotify support accounts. For each account, we used the search query \textit{`q=@account\_name\-\&f=tweets\-\&max\_position='} to crawl the tweets. The query parameter \textit{q} is set to a combination of the \textit{@}-symbol and the account name \textit{\{Netflixhelps, Snapchatsupport, SpotifyCares\}}. Thereby, we only consider tweets directly addressed to the support accounts (cf. Figure~\ref{fig:twitter-conversation}). We do not crawl tweets that solely use related hashtags (e.g., \textit{``Listen to my \#spotify playlist [...].''} or \textit{``Today, relaxed \#netflix sunday!''}). The type parameter \textit{f} is set to \textit{`tweets'} to receive all tweets addressed to the support accounts in temporal order, instead of only the top tweets as per default. The pagination parameter \textit{max\_position} is set to the identifier of the last tweet received, as the API returns a fixed amount of 20 tweets per request. For each tweet, we extracted the identifier (id), text, creation date, conversation id, reply flag, as well as the author's name and id.

Each tweet can result in a conversation which possibly contains responses written by the support team, by users facing similar issues, or by the reporting user (cf. Figure~\ref{fig:twitter-conversation}). To extract these responses, we additionally crawl each of the collected tweets status urls, following the pattern \textit{`https://twitter.com/user\_name/status/tweet\_id'}.

Table~\ref{tab:dataset} summarizes the \textit{crawled} dataset by the support accounts. 
The Netflix account (@Netflixhelps)~\cite{Weblink:115} was created the earliest in February 2009 and exists for about 10 years. For this account, we crawled 1,643,281 tweets by 385,935 users. These tweets result in 686,488 conversations ($\sim$2.4 tweets per conversation).
The Snapchat account (@Snapchatsupport)~\cite{Weblink:116} was created the latest in March 2014 and exists for about 5 years. We crawled 1,164,824 tweets by 422,643 users. These result in 612,645 conversations with about 1.9 tweets per conversation.
The Spotify account (@SpotifyCares)~\cite{Weblink:117} was created in February 2012 and exists for about 7 years. For this account, we crawled 2,446,864 tweets by 491,282 users, resulting in 892,441 conversations ($\sim$2.7 tweets/conversation).

The most frequented support account is Spotify with about 30 tweets and 11 conversations created per hour. Netflix and Snapchat are comparable with 19 tweets and 8-10 conversations per hour. The most active support team is Spotify with 1,256,465 (51.35\%) of the crawled tweets in all conversations created. Netflix created 752,951 (45.82\%) of the tweets, while Snapchat is least active with only 303,087 (26.02\%) tweets. 

Overall, the crawled dataset includes 5,245,969 tweets within 2,191,574 conversations, written by 1,299,860 users.

\begin{table}[t]
\renewcommand{\arraystretch}{1.3}
\caption{Key Figures of the Crawled Dataset.}
\label{tab:dataset}
\centering
\begin{tabularx}{\columnwidth}{lXXXX}
\toprule
 & \textbf{Netflix} & \textbf{Snapchat} & \textbf{Spotify} & \textbf{All} \\
\midrule
\# Tweets & 1,643,281 & 1,164,824 & 2,446,864 & 5,254,969 \\
\# Users & 385,935 & 422,643 & 491,282 & 1,299,860 \\
\# Conversations & 686,488 & 612,645 & 892,441 & 2,191,574 \\ \noalign{\smallskip}\cline{1-5}\noalign{\smallskip} 
Account created & 02/2009 & 03/2014 & 02/2012 & n/a \\
Tweets per hour & 18.71 & 18.73 & 29.53 & 22.32 \\
Conversations per hour & 7.82 & 9.85 & 10.77 & 9.48 \\
Tweets by support & 45.82\% & 26.02\% & 51.35\% & 41.06\% \\
\bottomrule
\end{tabularx}
\end{table}


\subsubsection{Truthset Creation Phase}

To be able to evaluate how well our approach extracts basic context items from tweets, we created a \textit{truthset} including labelled tweets of the Netflix, Snapchat, and Spotify support accounts.

Before creating the truthset, we pre-processed the tweets of the crawled dataset by removing conversations including non-English tweets using the LangID library~\cite{Weblink:101}. Then, we converted the tweet texts into lowercase, removed line breaks, double whitespaces, and mentions of support account names.

To create the truthset, we use the tool doccano~\cite{Weblink:102}, an open-source text annotation tool that can be used for, e.g., named entity recognition or sentiment analysis tasks. It can be deployed on a local or remote machine and offers rich functionality, such as user management. Using the tool, two human annotators performed a sequence labelling task by assigning the labels \textit{`Platform'}, \textit{`Device'}, \textit{`App Version'}, and \textit{`System Version'} to sequences within the tweets. 

We started from a random sample of conversations which resulted in truthsets nearly including no context items, being unusable to measure the performance of our approach. Thus, we changed the sampling strategy and searched for conversations including the keyword \textit{`App'}. The labelled context items were often referring to platforms such as desktops or smart TVs, which we do not consider in this paper. To select tweets including relevant context items, we only consider conversations containing the words \textit{`iOS'} or \textit{`Android'} in at least one of their tweets, even though this introduces the bias of more platforms being mentioned within the truthset. From the extracted conversations we randomly selected as much for each account to contain about 1,000 user tweets. We removed tweets written by the support teams as our approach is designed to extract context items from user feedback. Further, user tweets include more context items and are needed to determine how our approach performs on informal language, e.g., referencing the device \textit{`iPhone 6 Plus'} by the alternative spelling \textit{`iphone6+'}.

In case of disagreements between the two coders, a third annotator resolved the conflicts which resulted mainly from different sequence lengths due to including additional information, such as the device manufacturer or system architecture (e.g., \textit{`8.4.17'} vs. \textit{`8.4.17arm7'}). We calculated the inter-coder reliability using Cohen's Kappa on a scale of 0-1~\cite{cohen1960coefficient}. Per tweet of the truthset, we compare if the two coders agree or disagree that it includes context items. As suggested by Landis and Koch, we consider the ranges 0.61-0.80 as `substantial' and 0.81-1.00 as `almost perfect'~\cite{landis1977measurement}. The kappa agreement among the two coders is 0.933.

\begin{table}[t]
\renewcommand{\arraystretch}{1.3}
\caption{Key Figures of the Truthset.}
\label{tab:truthset}
\centering
\begin{tabularx}{\columnwidth}{lXXXl}
\toprule
 & \textbf{Netflix} & \textbf{Snapchat} & \textbf{Spotify} & \textbf{All} \\
\midrule
\# Conversations & 410 & 410 & 200 & 1,020 \\
\# User Tweets & 1,005 & 1,004 & 1,005 & 3,014 \\
\quad(incl. Context) & 379 (37.71\%) & 453 (45.12\%) & 284 (28.26\%) & 1,116 \\
\# Context Items & 546 & 736 & 558 & 1,840 \\
\midrule
\# Platform & 311 & 416 & 204 & 931 (50.60\%) \\
\# Device & 168 & 164 & 156 & 488 (26.52\%) \\
\# System Version & 56 & 130 & 109 & 295 (16.03\%) \\
\# App Version & 11 & 26 & 89 & 126 (6.85\%) \\
\bottomrule
\end{tabularx}
\end{table}

\begin{figure*}[t]
\centering
\includegraphics[width=\textwidth]{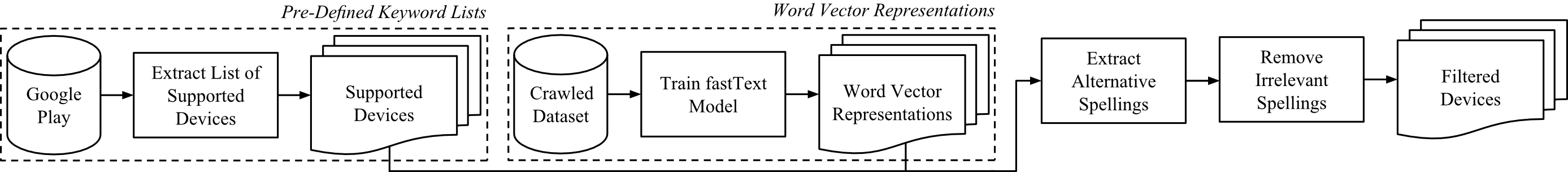}
\caption{Creation of the Filtered Android Device List using Pre-Defined Keyword Lists and Word Vector Representations.}
\label{fig:extraction-hitl}
\end{figure*}

Table~\ref{tab:truthset} summarizes the truthset. It consists of 1,020 conversations including 3,014 tweets, of which 1,005 are tweets from the Netflix support account, 1,004 tweets from Snapchat, and 1,005 tweets from Spotify.
Of these, 1,116 (37.03\%) tweets include context information. The tweets include an overall amount of 1,840 context items ($\sim$1.65 items per tweet), of which 931 (50.60\%) mention the platform, 488 (26.52\%) refer to the device, 295 (16.03\%) indicate the system version, and 126 (6.85\%) the app version.


\subsubsection{Data Analysis Phase}

In the data analysis, we answer how well basic context items can be automatically extracted from tweets. Therefore, we apply our approach to the truthset including labelled tweets of the Netflix, Snapchat, and Spotify support accounts. We measure the approach performance by comparing its output to the results of the human annotators. Then, we run summative analysis on the extracted information. Considering all support accounts, the approach achieved precisions from 81\% to 99\% and recalls from 86\% to 98\% for the different context item types. To support replication, our datasets and the analyses source code as Jupyter notebooks are publicly available on our website\footnote{https://mast.informatik.uni-hamburg.de/app-review-analysis/}.


\section{Context Extraction Approach}

We describe a simple approach that accurately extracts basic context information from unstructured, informal user feedback on mobile apps. We decided to consider the context items platform, device, app version, and system version, as we identified these four types to be frequently requested by support teams during a manual data exploration of 100 conversations. Moreover, researchers highlighted their importance for understanding and reproducing issue reports~\cite{bettenburg-fse-2008}. 

Our approach focuses on the Android and iOS platform. 
Both platforms cover 99.9\% of the mobile operating system market~\cite{Weblink:103}. The approach is designed to work with other platforms as well (e.g., desktop apps, smart TV apps), by exchanging its configuration files, i.e., the pre-defined keyword lists, without modifying the actual implementation.

We separate the description by the context item types and their strategies used for extraction.


\subsection{Platform and Device}

We crawl pre-defined keyword lists, including platform and device names, and generate word vector representations to handle informal writing frequently used in social media. Word vector similarities allow spelling mistakes and abbreviations of items included within the pre-defined lists to be determined. The lists and alternative spellings are used to create regular expressions that are applied to user feedback in order to extract context information. Figure~\ref{fig:extraction-hitl} summarizes our approach.

\subsubsection{Pre-Defined Keyword Lists} 

We crawled pre-defined lists of code names for the Android platform, as well as lists including device names for iOS and Android.
These lists are maintained by app store operators or user communities and updated regularly, e.g., with the release of new devices.

For the Android platform, 15 alternative code names exist, such as \textit{`Cupcake'}, which we extracted from a public list~\cite{Weblink:107}. For the iOS platform, no such alternative names exist.

For iOS devices we extracted 51 names, such as \textit{`iPhone 8 Plus'}~\cite{Weblink:104}. Since several users only refer to the product line, e.g., \textit{``[...] the error appears on my iPhone.''}, we extend the device list by the 5 product lines iPhone, iPad, iPod Touch, Apple TV, and Apple Watch, resulting in 56 iOS devices.

For Android devices the diversity is much higher. We crawled an official list from Google Play containing all 23,387 supported devices~\cite{Weblink:105}. The list includes four columns, listing the retail branding (e.g., \textit{`Samsung'}), marketing name (e.g., \textit{`Galaxy S9'}), device (e.g., \textit{`star2qlteue'}), and model (e.g., \textit{`SM-G965U1'}). We pre-process the list in five steps: 
We create a unique list of marketing names, as these possibly occur several times due to the same device being manufactured for different markets (e.g., European or Asian). The resulting list includes 15,392 devices. 
Then, we remove all marketing names shorter than 5 characters, such as \textit{`V'} or \textit{`Q7'}, resulting in 13,259 devices.
Further, we remove marketing names that are not mentioned within the collected tweets. We removed these, as word-vector models perform better on extracting similar words when a given input is included in the training data, while extracting alternative spellings for unseen words could negatively influence the results~\cite{bojanowski2017enriching}. It significantly reduced the number of devices to 1,324. This step needs to be repeated in fixed periods of time when new tweets are addressed to the support accounts.
As the list of marketing names also includes common words (e.g., \textit{`five'}, \textit{`go'}, or \textit{`plus'}), we used the natural language processing library spaCy~\cite{Weblink:106} to remove words that appear in the vocabulary of the included \texttt{en\_core\_web\_sm} model, trained on the CommonCrawl dataset. 
Thereby, we reduced the number of devices to 1,133. 

Until this point the processing of the keyword lists is fully automated. We decided to manually fine-tune the Android device list by removing remaining common names not included within the vocabulary of the CommonCrawl dataset (e.g., \textit{`horizon'}), while preserving more specific names (e.g., \textit{`galaxy s8'}), resulting in 896 Android devices. This step could possibly be automated with datasets of larger vocabulary sizes.

\subsubsection{Word Vector Representations} 

User feedback written in informal language might include alternative spellings of platform and device names, i.e., abbreviations or misspellings. For example, several users reference the Android code name \textit{`Lollipop'} as \textit{`lolipop'} or \textit{`lollypop'}. 

To enable our approach to also identify these cases, we create word vector representations using the fastText library~\cite{joulin2016bag}. Comparing vector distances allows to automatically identify similar words that frequently appear in the same context. A subset of these similar words are alternative spellings of the platform and device names included in our lists. 
We decided to use fastText over simpler methods, such as the Levenshtein distance, to also identify alternative spellings that vary significantly. For example, users often reference the \textit{`iPhone 6 Plus'} as \textit{`iphone6+'}, where the Levenshtein distance is 7. High edit distances would negatively impact the results by detecting, e.g., \textit{`one'} as alternative spelling to \textit{`iPhone 4'}, where the edit distance is 5.

To train the fastText model, we pre-process all 5,254,969 crawled tweet texts according to the truthset (i.e., we convert the tweet texts into lowercase, remove line breaks, double whitespaces, and mentions of support account names). 

\begin{algorithm}[t]
  \footnotesize
  \caption{Generates regular expression to extract context items from tweet texts, using pre-defined keyword lists and alternative spellings obtained from a trained word vector model.}
  \label{lst:keywords-wordvectors}
    
    \SetKwInOut{Input}{Input}
    \SetKwInOut{Output}{Output}

    \underline{def generateRegularExpression} $(k,t)$:
    \BlankLine
    
    \Input{Pre-processed keyword list $k$\newline Pre-processed tweets of app support account $t$}
    \Output{Regular expression to extract items in keyword list and their alternative spellings from text $r$}
    \BlankLine
    
    %
    \nonl\Comment{tokenize tweets (1)}\\
    
    import spacy\\
    nlp = spacy.load('en')
    \BlankLine
    
    tokenized\_tweets = []\\
    \For{tweet in t}{
        tokens = []\\
        \For{token in nlp(tweet)}{
            \If{not (token.is\_punct or token.is\_space)}{
                tokens.append(tokens)\\
            }
        }
        tokenized\_tweets.append(tokens)\\
    }
    
    %
    \nonl\Comment{train word vector model (2)}\\
    
    from gensim.models.fasttext import FastText\\
    ft = FastText(size=300, window=5, min\_count=5)\\
    ft.build\_vocab(sentences=tokenized\_tweets)\\
    ft.train(sentences=tokenized\_tweets, epochs=10)
    \BlankLine
    
    %
    \nonl\Comment{extract alternative spellings (3)}\\
    
    alternative\_spellings = []\\
    \For{keyword in k} {
      similar\_words = ft.wv.most\_similar(keyword, topn=10)\\
      \For{word, similarity in similar\_words} {
        \If{similarity $>=$ threshold} {
          alternative\_spellings.append(word)
        }
      }
    }
    \BlankLine
    
    %
    \nonl\Comment{generate regular expression (4)}\\
    
    r = '$|$'.join(k + alternative\_spellings)\\
    \Return{r}
\end{algorithm}

\begin{figure}[b]
\centering
\includegraphics[width=.85\columnwidth]{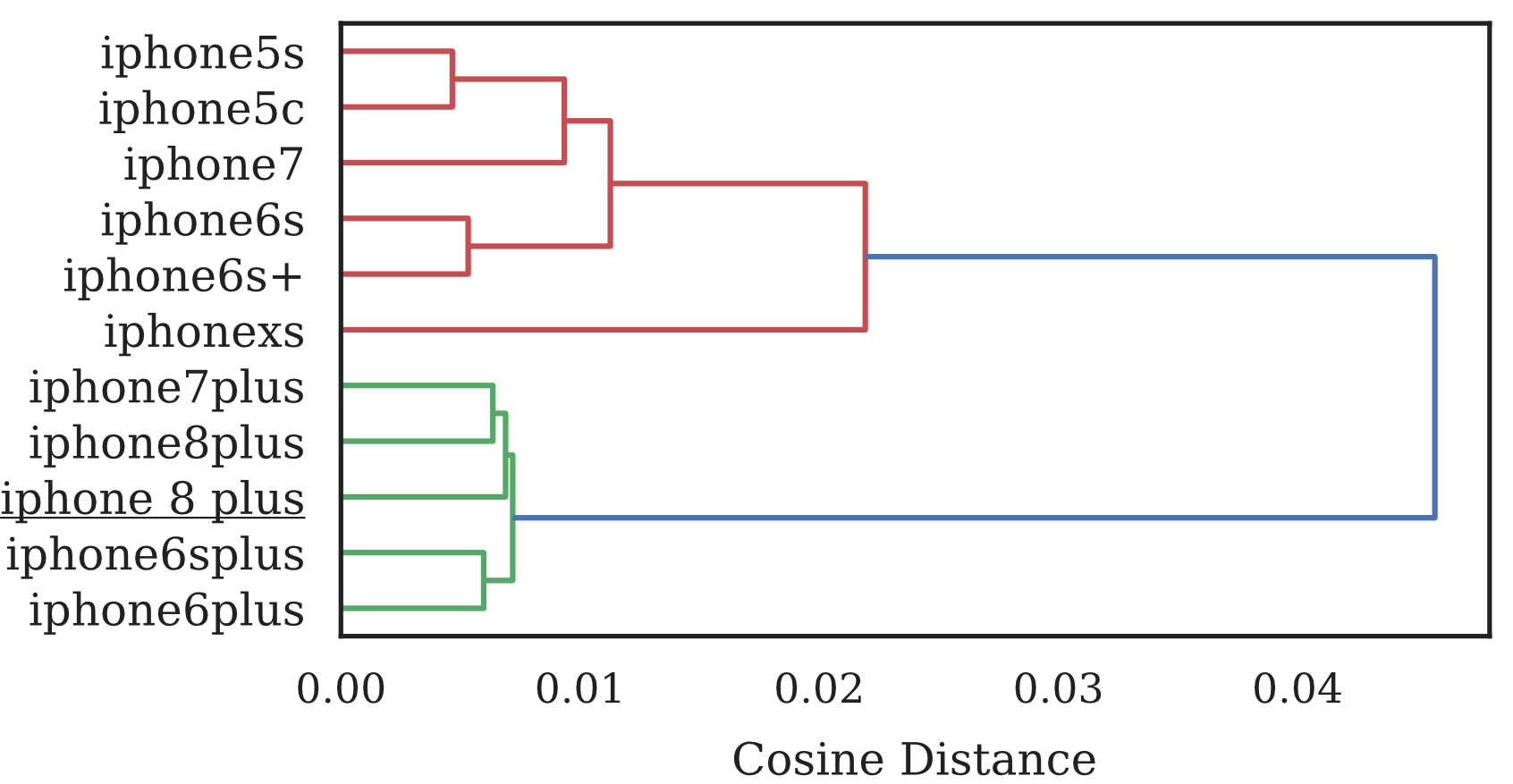}
\caption{Ten most similar Words to `iPhone 8 Plus' (i.e., Misspellings or Abbreviations), identified using the trained fastText Model.}
\label{fig:wordvectors-similar}
\end{figure}

Algorithm~\ref{lst:keywords-wordvectors} lists the extraction of similar spellings for given keywords using word vector representations as pseudocode. It takes the pre-processed tweets and a keyword list (i.e., including the iOS devices names or Android platform code names) as input. The algorithm can be separated into four parts: First it tokenizes the tweets and removes non-informative tokens (line 2-13), then it trains the word vector model using the tweets (line 14-17), afterwards it obtains alternative spellings for each given keyword from the word vector model (line 18-26), finally it generates a regular expression of the original keywords and their alternative spellings (line 27-28).
In the following, we explain each part separately:

\begin{enumerate}[label=\textit{(\arabic*)}, leftmargin=*]
  \item \textit{Tokenize Tweets.} We begin by tokenizing each tweet. We remove non-informative tokens including punctuation and spaces using spaCy's~\cite{Weblink:112} built-in functionality.

  \item \textit{Train Word Vector Model.} For the actual training of the model, we use Gensim \cite{rehurek_lrec} as suggested by spaCy. We use the default configuration and set the word vector size to 300, the minimum occurrences of words to 5, the window size to 5, and perform the training in 10 epochs. Our trained model has a vocabulary size of 149,889 words.
  
  \item \textit{Extract Alternative Spellings.} We extracted similar words per platform and device name included in our lists. Figure~\ref{fig:wordvectors-similar} shows the 10 most similar words and their distances for the device \textit{`iPhone 8 Plus'}, clustered in a dendrogram. The most similar word is \textit{`iphone8plus'} (cosine distance of 0.047), followed by \textit{`iphone6plus'} (0.051). Other device names, such as \textit{`zenphone'}, are also extracted but have a higher cosine distance, in this case 0.328. These can be automatically filtered by setting a fixed threshold for the similarity (line 9), we used a threshold of 0.2. 
For the Android code names we extracted 14 unique alternative spellings. 
For iOS devices we extracted 44 alternative spellings and 392 for Android devices. 

  \item \textit{Generate Regular Expression.} Per list, we combine the given keywords (e.g., devices) and their alternative spellings into a single regular expression using the `OR' operator, e.g., \textit{`iPhone XR$\rvert$...$\rvert$iPhone 7'}. Later, we apply the Python functionality \texttt{re.search(pattern, string)}~\cite{Weblink:110} to the user feedback. As users also include multiple devices, such as \textit{``[...] the error occurs on my iPhone 6 and iPad Mini.''}, we modify the function to return the locations of all matches within a given input.
\end{enumerate}

\subsubsection{Manual Fine-Tuning of Results} 

The proposed approach to extract context items using pre-defined keyword lists and word vector representations can be run completely automated. Whenever, e.g., new devices are released, the keyword lists are updated by the app store operators or user communities. These, as well as the updated tweets dataset, including the most recent tweets of an app support account which possibly contain alternative spellings of new device names, need to be regularly provided as input to Algorithm~\ref{lst:keywords-wordvectors} to update the regular expression used to extract the context items. To fine-tune the results, we invested manual effort at two points.

First, when pre-processing the keyword lists to extract alternative spellings using word vectors, we manually removed device names solely consisting of common words (e.g., \textit{`horizon'}) that could not be automatically removed. From the original 1,133 devices, we thereby removed 237 devices. This manual effort latest for about two hours. It needs to be repeated regularly, e.g., when new devices are added to the pre-defined keyword lists. However, in these cases the effort is significantly lower since only single device names need to be processed instead of all supported devices since the release of Google Play about 10 years ago.

Second, we decided to manually filter alternative spellings for Android code names. We found that these include words (such as \textit{`bake'}) that are unrelated to the inputs in the context of software engineering (such as the Android code name \textit{`pie'}). Thereby, we removed 6 out of the 14 alternative spellings.
Similar, we processed the alternative spellings for Android and iOS devices. For Android, we removed 326 spellings out of 392 alternative spellings. For iOS, we removed 22 of the 44 alternative spellings. We assume that more Android devices names have been removed, since the device names are very diverse and not as often included within the tweets which causes word vector models to suggest similar words that are not as closely related, compared to the names of iOS devices. This manual step lasted under an hour and is -- as for the first step --   significantly faster for future updates.


\subsection{App and System Version}

To extract app and system versions from user feedback, we crawled pre-defined keyword lists and created text patterns. The keyword lists include the released app and system versions. We collected 107 system versions for iOS and 59 for Android. Concerning the app versions~\cite{Weblink:108, Weblink:109}, we extracted 224 iOS and 133 Android versions for Netflix, 248 iOS and 346 Android versions for Snapchat, as well as 169 iOS and 165 Android versions for Spotify.

We tokenize the user feedback with spaCy. Then, we pre-process all tokens by removing leading characters before digits, such as \textit{`v8.4.17'}. If the leading characters equal a platform (e.g., \textit{`iOS12'}), we split the token to keep the platform. We also remove trailing characters often referring to system architectures, as such as \textit{`8.1.13arm7'}.
This might be a limitation that has to be adapted for other platforms. Versions for the platforms considered in our study cannot be named with leading or trailing characters (e.g., \textit{`A1.0'} or \textit{`1.1a'}).

By manually comparing the collected versions to those mentioned in the user feedback, we identified two challenges. 
First, the collected versions have intersections. For example, version 7.1.2 exists for the Netflix iOS app, as well as for both the iOS and Android operating system. Therefore, we cannot directly associate it with, e.g., the Android operating system. The intersections highly vary, the Netflix iOS app shares only 8 (3.57\%) out of 224 versions with its Android app. In contrast, for Snapchat the app versions are much more similar with an intersection of 32.66\% between iOS and Android. A relatively large overlap also exists for Android system versions and versions of the Netflix iOS app (27.12\%), as well as the iOS operating system (42.37\%). 
The second challenge are users reporting more detailed app versions (e.g., \textit{`8.0.1.785'}) than included in the public lists. In this example, the user refers to the Snapchat Android app where the list only includes the version \textit{`8.0.1'}, missing the subversion \textit{`.785'}.

\begin{figure}[b]
\centering
\includegraphics[width=.75\columnwidth]{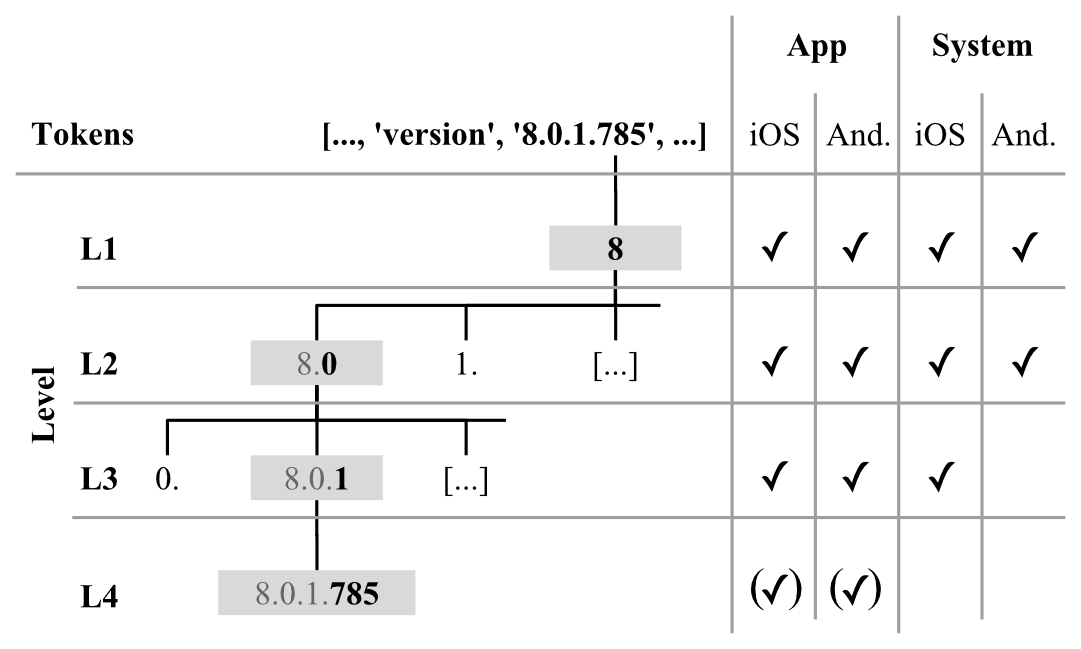}
\caption{Version Matcher processing the Input `version 8.0.1.785'.}
\label{fig:version-matcher}
\end{figure}

\begin{algorithm}
\footnotesize
    \SetKwInOut{Input}{Input}
    \SetKwInOut{Output}{Output}

    \underline{def matchVersion} $(c,l_{sys-ios}, l_{sys-and}, l_{app-ios}, l_{app-and})$:
    \BlankLine
    
    \Input{Conversation with multiple tweets (or single tweet) $c$
      \newline Lists of versions $l_{x}$
      \newline Platform (if known) $p$
      \newline Device (if known) $d$}
    \Output{List of extracted and labelled versions $v$}
    \BlankLine
    
    %
    \nonl\Comment{generate version tree (1)}\\
    version\_tree = Tree()\\ 
    \For{l in ($l_{sys-ios}, l_{sys-and}, l_{app-ios}, l_{app-and}$)} {
      \For{version, label in l} {
        version\_tree.add(version, label)
      }
    }
    \BlankLine
    
    %
    \nonl\Comment{process conversation or tweet (2)}\\
    
    extracted\_versions = []\\
    \For{tweet in c}{
      \For{token in tweet}{
          potential\_matches = version\_tree.match(token, previous\_token)\\
          extracted\_versions.append(token, potential\_matches)\\
          previous\_token = token
      }
    }
    \BlankLine
    
    %
    \nonl\Comment{\textit{optional:} resolve conflicts (3)}\\
    
    \For{v, potential\_matches in extracted\_versions}{
      \If{count(potential\_matches) $>$ 2}{
        \uIf{(p.is\_Android() or d.is\_Android()) and (!p.is\_iOS() and !d.is\_iOS())}{
          potential\_matches.remove\_iOS()\\
        }
        \uElseIf{(p.is\_iOS() or d.is\_iOS()) and (p.is\_Android() or d.is\_Android())}{
          potential\_matches.remove\_Android()\\
        }
      }
    }
    
    \Return{extracted\_versions}
    \BlankLine

    \caption{Generates version tree to extract app and system versions from text (simplified).}
    \label{lst:version-matcher}
\end{algorithm}

We implement a version matcher to handle these challenges, shown in a  simplified manner in Algorithm~\ref{lst:version-matcher}. As input, it takes a conversation consisting of multiple tweets or single tweets, as well as the version lists. Further, the previously extracted platform and device can be provided as input, if exists. The algorithm can be separated into three steps: First, it generates a version tree (line 2-7). Then, it processes conversations or single tweets to extract included versions (line 8-15). Optionally, it resolves existing conflicts (line 16-24).
In the following, we explain these steps separately:

\begin{enumerate}[label=\textit{(\arabic*)}, leftmargin=*]
  \item \textit{Generate Version Tree.} Each version list and respectively their included versions are processed to create a version tree. If the list of app versions for iOS, e.g., includes the version \textit{`8.0.1'}, the version is split by its subversions and three leaves are added to the tree (i.e., \textit{`8'}, \textit{`8.0'}, and \textit{`8.0.1'}). Each leave is marked as an iOS app version. If the list of system version for Android includes the version 8.0, no leaves are added to the version tree. Instead the existing leaves \textit{`8'} and \textit{`8.0'} are marked as versions of both the iOS app and Android system.
  
  \item \textit{Process Conversation or Tweet.} The matcher takes each token including a number and respectively its previous token as input. Figure~\ref{fig:version-matcher} shows the matcher traversing the version tree on separate levels (L1-L4) to process the input \textit{`version 8.0.1.785'}. The subversion \textit{`.785'} (L4) is not included in our crawled lists. Therefore, the closest version, i.e., \textit{`8.0.1'}, of the previous level (L3) is selected. This version exists for both the Snapchat iOS and Android app, as well as the iOS system. As noted previously, not all app versions were included in the pre-defined lists, however we know that the collected list of system versions is complete. For this reason, the iOS operating system is removed as potential match (cf. Figure~\ref{fig:version-matcher}). If multiple system versions would remain, the matcher would process the previous token. If this token equals \textit{`iOS'} or \textit{`Android'}, the matcher flags this respectively as iOS or Android system version. This is especially relevant for shorter versions, such as \textit{`8'} or \textit{`8.0'}, where more potential matches exists. Since several possible matches remain, i.e., the version could refer to the iOS or Android app, it is conflicted and will be processed in the next phase.
  
  \item \textit{Resolve Conflicts (optional).} If conflicts remain, potential version matches are assessed in their overall conversation context. Another feedback in the conversation might include additional context items, e.g., as a device either for Android or iOS, which helps to determine which platform the version is referring to. In the example conversation, the user previously wrote \textit{``The error occurs on my HTC One with Android installed.''}. As this feedback includes the Android platform and device, both context items are provided as input to Algorithm~\ref{lst:version-matcher} (parameters $p$ and $d$). If one of these relates to Android and none to iOS (line 18), the conflict is resolved by marking the version as Android app. A limitation of our approach are tweets, such as \textit{``It worked with my Galaxy S5, but is not working with my new Galaxy S6''}. In this case, the conflict would not be resolved and both devices would be extracted. We consider this as beneficial, as knowing that the error occurs on one device but not the other might help developers. However, automatically highlighting on which of the devices the reported issue does not occur requires more complex natural language processing approaches, which we do not consider as focus of our study.
\end{enumerate}


\section{Evaluation Results}

We evaluated the performance of our approach to extract basic context items (including the platform, device, app version, and system version)  by comparing its results to the manually labelled truthset. Our truthset contains 3,014 tweets of the Netflix, Snapchat, and Spotify support accounts. Of these, 1,116 (37.03\%) tweets include an overall amount of 1,840 context items (cf. Section~II). 

Table~\ref{tab:tool-performance} summarizes the results per context item and support account. The table shows the  number of corresponding items in the truthset, the number of true positives (i.e., correctly identified context items), false positives (incorrect identified items, such as \textit{`galaxy s8'} instead of \textit{`galaxy s8 plus'}), false negatives (no items extracted although present in tweet), and true negatives (no items detected, where no items are present). Based on these, the approach's precision and recall is calculated. The precision indicates how many of the extracted items are correctly identified. The recall summarizes how many of all items included in the truthset were extracted. The table further combines the results per context item type for different apps by calculating their average. For different types, the precision varies from 81\% to 99\%, and the recall from 86\% to 98\%.


\subsubsection{Platform} 

The platform is most frequently provided within the truthset, with 931 (50.60\%) out of 1,840 context items. Of all platform mentions, our approach extracted 910 correctly (true positives) and missed to extract the remaining 21 (false negatives). The absence of the platform was correctly detected in 2,117 tweets (true negatives). 
For this context type, we refrain from reporting the precision. The truthset is biased, since we sampled for conversations that include the words \textit{`Android'} or \textit{`iOS'} in one of the tweets, to increase the amount of labelled context items. This ratio is not representative for the whole dataset. However, this is the least complex context item to extract and alternative platform code names, such as \textit{`Gingerbread'}, were successfully extracted. 

The recall for the platform is 98\%. False negatives result from alternative spellings of Android code names that are not used frequently. For these, additional tweets need to be collected to train the fastText model or its minimum occurrences of words has to be tuned to increase the vocabulary size.


\subsubsection{Device} 

Users within the truthset report 488 (26.52\%) context items referencing a device. Our approach identified 386 true positives, 38 false positives, 64 false negatives, and 2,544 true negatives. For this type, the approach achieved a precision of 91\% and a recall of 86\%.

The detected false positives, e.g., include the device \textit{`Galaxy S8'} within the tweet \textit{``[...] android version 8.0.0 galaxy s8 plus for t-mobile''}. Here the user used the device name \textit{`Galaxy S8 Plus'} instead of \textit{`Galaxy S8+'} included in the list.
False negatives primarily consist of shortened device names, such as the \textit{`Samsung Galaxy S5'} mentioned as \textit{`s5'} within the tweet \textit{``[...] worked fine on my iphone and laptop, just not on my s5.''}. Other short device names include \textit{`g4'} (referencing tweet: \textit{``android 7.0 on a moto g4 [...]''}), \textit{`1610'} (\textit{``[...] vivo 1610 android : 6.0.1 [...]''}), \textit{`s8'} (\textit{``s8 running android 7.0 [...]''}), and \textit{`s9+'} (\textit{``[...] phone is a samsung s9+, so android [...]''}). Part of these devices has been removed while pre-processing the device lists by filtering short device names, such as the \textit{`1610'} (cf. Section~III).
The approach results might improve by adding short devices names in combinations with their manufacturer (e.g., \textit{`Vivo 1610'}) to the device lists. Short names have been previously excluded to reduce false positives.

\begin{table}[]
\renewcommand{\arraystretch}{1.3}
\caption{Performance of the Approach Compared to Truthset. (T/FP = True/False Positive, F/TN = False/True Negative)}
\label{tab:tool-performance}
\centering
\begin{tabularx}{\columnwidth}{lllXXXlXX}
\toprule
\textbf{Type} & \textbf{Account} & \rotatebox{90}{\textbf{\# Items}} & \rotatebox{90}{\textbf{TP}} & \rotatebox{90}{\textbf{FP}} & \rotatebox{90}{\textbf{FN}} & \rotatebox{90}{\textbf{TN}} & \rotatebox{90}{\textbf{Precision}} & \rotatebox{90}{\textbf{Recall}} \\ 
\midrule
\multirow{4}{*}{\rotatebox[origin=c]{90}{\textit{Platform}}} 
 & Netflix & 311 & 303 & 0 & 8 & 701 & n/a & 0.97 \\
 & Snapchat & 416 & 403 & 0 & 13 & 615 & n/a & 0.97 \\
 & Spotify & 204 & 204 & 0 & 0 & 801 & n/a & 1.00 \\
 & \textbf{Combined} & \textbf{931} & \textbf{910} & \textbf{0} & \textbf{21} & \textbf{2,117} & \textbf{n/a} & \textbf{0.98} \\
\midrule
\multirow{4}{*}{\rotatebox[origin=c]{90}{\textit{Device}}} 
 & Netflix & 168 & 140 & 8 & 20 & 845 & 0.95 & 0.88 \\
 & Snapchat & 164 & 130 & 12 & 22 & 840 & 0.92 & 0.86 \\
 & Spotify & 156 & 116 & 18 & 22 & 859 & 0.87 & 0.84 \\
 & \textbf{Combined} & \textbf{488} & \textbf{386} & \textbf{38} & \textbf{64} & \textbf{2,544} & \textbf{0.91} & \textbf{0.86} \\
\midrule
\multirow{4}{*}{\rotatebox[origin=c]{90}{\textit{App Version}}} 
 & Netflix & 11 & 7 & 3 & 1 & 994 & 0.70 & 0.88 \\
 & Snapchat & 26 & 17 & 9 & 0 & 977 & 0.65 & 1.00 \\
 & Spotify & 89 & 74 & 11 & 4 & 918 & 0.87 & 0.95 \\
 & \textbf{Combined} & \textbf{126} & \textbf{98} & \textbf{23} & \textbf{5} & \textbf{2,889} & \textbf{0.81} & \textbf{0.95} \\
\midrule
\multirow{4}{*}{\rotatebox[origin=c]{90}{\textit{System Ver.}}} 
 & Netflix & 56 & 47 & 1 & 8 & 948 & 0.98 & 0.85 \\
 & Snapchat & 130 & 116 & 0 & 14 & 876 & 1.00 & 0.89 \\
 & Spotify & 109 & 88 & 2 & 19 & 898 & 0.98 & 0.82 \\
 & \textbf{Combined} & \textbf{295} & \textbf{251} & \textbf{3} & \textbf{41} & \textbf{2,722} & \textbf{0.99} & \textbf{0.86} \\
\bottomrule
\end{tabularx}
\end{table}


\subsubsection{App Version} 

The truthset includes 126 items (6.85\%) reporting the app version. Our approach detected 98 true positives, 23 false positives, 5 false negatives, and 2,889 true negatives. The approach precision is 81\% and recall 95\%.

The detected false positives, e.g., include the version \textit{`0.9.0.133'} appearing in the tweet \textit{``version 0.6.2.64 on the phone and think its 0.9.0.133 on the desktop''} from the Spotify dataset. In the tweet the user also refers to the desktop version, however version \textit{`0.9.0'} also exists for the Spotify iOS app. Other false positives include the version `3.0' and `4.0' which are detected as app version while referring to system versions, e.g., in the tweets \textit{``[...] cant connect with spotify on android since 3.0 update [...]''} or \textit{``[...] i try to sync my ipod shuffle 4.0. any help? [...]''}. The false positives mainly result from intersections between app and system versions of different platforms (cf. Section~III, B.). The intersections are the highest for Snapchat, resulting in a precision of 65\%.
False negatives are rare with only 5 occurrences and include versions not correctly separated by dots, such as `2.07' instead of `2.0.7'.


\subsubsection{System Version} 

Users report 295 context items (16.03\%) including a system version. The approach identifies 251 true positives, 3 false positives, 41 false negatives, and 2,722 true negatives. For this context type, our approach achieved a precision of 99\% and recall of 86\%.

False negatives mainly result from the version matcher only taking into consideration a potential version's previous token to decide if this refers to the system. This applies, e.g., for the tweet \textit{``on android (8.1 pixel xl) [...]''} where the previous token is `(', as well as for the tweet \textit{``[...] android v 7.0 on spotify 8.4.39.673 armv7''} where \textit{`v'} is the previous token. For these matching if one of the two previous tokens equals \textit{`iOS'} or \textit{`Android'} would improve the results. Other, false positives result from misspellings of users, such as \textit{`iso7'} instead of \textit{`ios7'}. Also, more complex patterns which one user applies to report the system and app versions of multiple devices are not considered by the version matcher, as in the tweet \textit{``it also happens on my ipad (ios 10.1.1, spotify 6.8.0) and my wife's ipad (10.1.1/6.8.0) and iphone (10.1.1/6.8.0)''}.


Only 15 tweets within the truthset were marked as conflicted. An example tweet reports a Spotify app version, that exists for both iOS and Android \textit{``i'm on 8.4.74. doesnt bother me too much... just thought i'd report it''}. For conflict resolution, other tweets within the conversation are analyzed, such as \textit{``just so you know... the toast keeps going out of sync with what's actually playing at the moment. using a pixel 2 on android 9.''}. In this tweets the user reports both the Android platform and Android device. The conflict is resolved by marking the version as Android app version. All 15 conflicts were resolved, as we analyzed only completed conversations.


\section{Discussion}

\subsection{Implications}
Many software vendors, including Netflix, Snapchat, and Spotify, have recognized the advantages of gathering, analyzing, and reacting to user feedback provided via social media. About one third of the bugs reported in issue trackers can be discovered earlier by analyzing tweets~\cite{Mezouar:2018:TUB:3231288.3231333}. Speed is certainly a major advantage of social media-like feedback channels. Compared to reviews in app stores, the conversational nature of Twitter allows additional features and bugs to be identified. However, for the reported issues to be actionable to developers, basic context information, such as the utilized app version or device, needs to be included. 

Our research shows that support teams themselves are very active, providing $\sim$40\% of the tweets within the crawled dataset, in many cases to clarify missing context items. Spotify and other vendors also initiated local support teams, such as \textit{@SpotifyCaresSE} for Swedish users, with multiple involved persons~\cite{Weblink:113}. Smaller teams receiving a large amount of feedback might not be able to afford such a large investment.

This paper introduced a simple unsupervised approach that identifies the presence of and extracts context items from tweets. The results of our approach can be primarily used to filter actionable issues, i.e., conversations in which basic context items are present. When present, tweet texts and included context items can be used to auto-populate issue trackers with structured information \cite{bettenburg-fse-2008, bettenburg-msr-2008}.

Other conversations including only part of the basic context items might be non-actionable to developers. These can be automatically identified using our approach, as well as the exact information missing. When continuously applied to tweets, the output of our approach can be used, e.g., by a chatbot to immediately request missing context items from users by responding to conversations, e.g., \textit{``Can you tell us the device you are using?''}. Both measures help reduce the manual efforts of support teams on social media.

Besides support teams and developers, our approach can also assist users. Users -- often lacking software engineering and issue tracking knowledge -- might be unaware of the importance of context information and therefore simply exclude them from the feedback. In a first step,  users can be made aware about the importance of context items. While composing a tweet, this can continuously be analyzed to detect if a bug is reported, a feature is requested, or if the user might simply provide praise \cite{7320414}. When reporting a bug, a message can be shown to the user that the issue reported might only be actionable to developers when including context information. The context items that the user already included while writing the tweet, can be identified using our approach, and missing context items can even be suggested in-situ~\cite{Maalej:2011ku}.


\subsection{Limitations and Threats to Validity}

The support accounts on Twitter which we selected for our study are all of popular apps that appear within the top 25 charts of the Apple App Store. To improve the generalizability of our results, further support accounts for apps of different popularity (i.e.~receiving different amounts of feedback) should be considered in future studies. Also, further studies need to be carried out to determine if the type of selected apps might correlate with the amount of non-/technical users and possibly the amount of context items exchanged.

To create the truthset, we extracted only conversations including at least one of the keywords \textit{`Android'} or \textit{`iOS'}. Without this step the amount of context items in the truthset would have been too small. As these keywords are also detected as platform context, the percentage of context items reporting the platform might be not representative for the whole dataset. We also tried more general keywords, such as \textit{`App'} or no keywords at all but the extracted tweets included much less context items or context information related to platforms which we do not consider in our paper, such as Windows, Mac, or Linux. Further studies need to determine how our approach performs when considering all platforms supported by an app. Nevertheless, other identifiers for the platform, such as code names for Android versions (e.g., \textit{`Froyo'}) could successfully be extracted by our approach.

The pre-defined keyword lists extracted for the platform, device, app version, and system version certainly influence our results. These need to be updated regularly for platforms and apps our approach is applied to. The results for the app and system version are negatively affected if the app versions are equal or similar to the system versions. To improve this circumstance, we consider the previous token to detect if a potential version refers to the Android or iOS system version. Further studies should not only consider the previous token but use different window sizes of tokens before and possibly after a potential version to increase the accuracy of the approach.

Finally, to improve our results we trained a fastText model on all collected tweets. We extracted similar spellings for the platform and device names of our pre-defined lists. We manually identified relevant alternative spellings from the extracted similar words, such as \textit{`iphone6+'} for the input \textit{`iPhone 6 Plus'}. This might introduce errors. Further tweets need to be collected to train the fastText model and determine if it provides more similar words to given inputs, such as device names. Then, this manual step can possibly be automated by only using a fixed threshold for the cosine distance.


\section{Related Work}


Users provide an increasing amount of feedback on software products, e.g., in form of app reviews. Studies repeatedly showed that a significant amount of app reviews include information which is potentially useful to developers, such as shortcomings of app features, improvement requests, and bug reports \cite{6224306, 6636712, 7320414, Pagano:2013:UIS:2486788.2486920}. However, studies also found that that negative feedback often misses useful details as context information \cite{7765038, 6636712}. To solve this issue, Maalej et al.~\cite{7325177} suggested to improve feedback quality by automatically collecting context information. Many software vendors offer such built-in options and automatically attach relevant context information to reported issues. Our approach complement this direction by focusing on users who only exchange text information.

Research found that especially non-technical end-users are more likely to express their opinions on social networks, such as Twitter \cite{Mezouar:2018:TUB:3231288.3231333}. 
Several studies have identified Twitter as an important source for crowd-based requirements engineering and software evolution \cite{Nayebi:2018:EMSE,8048893, Mezouar:2018:TUB:3231288.3231333}. Similar to app reviews, tweets contain important information, such as feature requests or bug reports.
By performing a survey with software engineering practitioners and researchers Guzman et al.~\cite{7972758} underlined the need for automatic analysis techniques to, e.g., summarize, classify, and prioritize tweets. The authors highlight that a manual analysis of the tweets is unfeasible due to its quantity, unstructured nature, and varying quality.
Nayebi et al.~\cite{Nayebi:2018:EMSE} found that tweets provide additional requirements-related information. Compared to app reviews, by mining tweets the authors extracted $\sim$22\% additional feature requests and $\sim$13\% additional bug reports.
Other authors have used tweets to crowdsource app features~\cite{7961668}, to support release decisions~\cite{8491124}, to categorize and summarize technical information included in tweets~\cite{8048885}, or to rank the reported issues~\cite{8048886}. These studies enforce the relevance of our approach.

Hassan et al. \cite{Hassan2018} studied the dialogue between users and developers in Google Play. Within their dataset of 4.5 million app reviews, the authors identified and analyzed 126,686 responses by developers. The authors did not focus on context items but on conversations in general to show that reviews in app stores are not static. Compared to our results, developers in the dataset of Hassan et al. (i.e., on Google Play) are much less active with about 3\% of all reviews being from developers. Within our dataset developers provided 26\% to 51\% of all tweets. Similarly, Bailey et al.~\cite{bailey2019examining} studied the dialogue between users and developers. The authors focused on the Apple App Store and found that developers reply to reviews for about one fifth of the analyzed apps. Most discussed topics by developers are log-in issues, feature requests, and crashes, possibly also by clarifying missing context information.


\section{Conclusion}

Despite built-in options to report issues in a structured manner, users continue to share a large amount of unstructured, informal feedback on software products via social media. This feedback contains information of relevance to development teams, such as bug reports or feature requests. Support teams engage in effortful conversations with users to clarify missing context information -- for popular apps such as Spotify or Netflix in about 10 parallel conversations per hour.

We introduced a simple unsupervised approach to identify and extract basic context items from user feedback, including the affected platform, device, app- and system version. Evaluated against a manually labelled truthset of 3014 tweets, our approach achieved precisions from 81\% to 99\% and recalls from 86\% to 98\% for the different context item types.
Our approach can assist support teams to identify and separate reported issues into non-/actionable. Actionable issues can be used to auto-populate issue trackers with structured information. Non-actionable issues can be automatically clarified, e.g., by chatbots requesting the missing context items from users.


\section*{Acknowledgment}

This research was partially supported by the EU Horizon 2020 project OpenReq under grant agreement no. 732463.


\bibliographystyle{IEEEtran}
\bibliography{refs}

\end{document}